\documentstyle[12pt,graphics,epsfig]{article}
\begin{document}
%\draft
%\twocolumn
%\wideabs{
\title{The theory of heating of the quantum ground state of trapped 
ions}
\author{Daniel F. V. James,\\
\small{ Theoretical Division T-4, Mail Stop B-268,}\\ 
\small{Los Alamos National Laboratory, Los Alamos, NM 87545, USA} }
\date{\today}
\maketitle
%%%%%%%%%%%%%%%%%%%%%%%%%%%%%%%%%%%%%%%%%%%%%%%%%%%%%%%%%%%%%%%%%%%%%%%%%%%%%
% Abstract
\begin{abstract}
Using a displacement operator formalism, I analyse the
depopulation of the vibrational ground state of trapped ions.
Two heating times, one characterizing short time behaviour, 
the other long time behaviour are found.  The short time
behaviour is analyzed both for single and multiple ions, and a 
formula for the relative heating rates of different modes 
is derived. The possibility of correction of heating via the 
quantum Zeno effect, and the exploitation of the suppression of 
heating of higher modes to reduce errors in quantum computation 
is considered.
\end{abstract}
%%%%%%%%%%%%%%%%%%%%%%%%%%%%%%%%%%%%%%%%%%%%%%%%%%%%%%%%%%%%%%%%%%%%%%%%%%%%%
\bigskip
LAUR 98-654\\
%%%%%%%%%%%%%%%%%%%%%%%%%%%%%%%%%%%%%%%%%%%%%%%%%%%%%%%%%%%%%%%%%%%%%%%%%%%%%
%\pacs{PACS numbers:42.50.Vk, 32.80.Pj, 03.67.Lx, 42.25.Kb}}\narrowtext
%%%%%%%%%%%%%%%%%%%%%%%%%%%%%%%%%%%%%%%%%%%%%%%%%%%%%%%%%%%%%%%%%%%%%%%%%%%%%
%\newpage
%%%%%%%%%%%%%%%%%%%%%%%%%%%%%%%%%%%%%%%%%%%%%%%%%%%%%%%%%%%%%%%%%%%%%%%%%%%%%
%Main Text
Individual or multiple ions can be confined in a radio-frequency Paul
trap and using sophisticated laser techniques 
cooled to the quantum mechanical ground state 
\cite{Diedrich:89,Monroe:95a}.  Such
systems allow experimental preparation and measurement of 
non-classical motion states of the ions, and are therefore of great
current interest in physics \cite{Leibfried:97}.  
Furthermore important technological
applications for such systems, such as the practical implementation
of quantum computation, have recently attracted considerable attention
\cite{Cirac:95,Monroe:95b,Steane:97,James:98} .
Quantum information can be stored in the internal quantum
states of the ions (which constitute the quantum bits, or
``qubits'' of the computer), and, using ultra narrow bandwidth lasers,
quantum gate operations can be realized between pairs of
qubits using quantum states of the collective motion of
the ions in the harmonic confining potential as a quantum
information bus.  If this bus were to become degraded by heating
information would be lost, and so it is of great importance
to maintain the ions in their motional ground state as long
as possible.
Of the many practical roadblocks standing in the way of
success in realization of an ion trap quantum computer, one
of the most important is the very fragile nature of this motional
ground state,
due to interactions with various ambient electromagnetic fields.
In this letter I present a theoretical analysis of the depopulation
of the motional ground state of ions due to such fields, which can
be loosely characterized as heating (although it should be stressed
that in this analysis
relaxation to a thermal distribution is {\em not} 
considered).

Various analyses of decoherence mechanisms in ion trap quantum
computers have been carried out
\cite{Garg:96,Hughes:96,Plenio:97,James:97,Wineland:97}.  
Such effects as spontaneous
emission from the internal states 
\cite{Hughes:96,Plenio:97,James:97}, or dephasing of the
internal states due to the ions' zero-point motion
\cite{Garg:96}, both of which effects
can degrade or destroy quantum information, have been considered.
Also the relaxation of an ion to a thermal state has been
considered using a perturbation method, valid for describing
long time behavior \cite {Lamoreaux:97}; other work has been
done on the effects of laser amplitude and phase stability
\cite{Schneider:98,Knight}.  

In this letter I adopt a rather different approach than that
normally used in analysis of decoherence problems.
When the electromagnetic fields causing the heating can be treated
classically, the equations of motion reduce to a solvable system,
namely a simple harmonic oscillator driven by a classical field.
One can thus dispense with the
density matrix/master equation formalism, using instead the 
interaction picture to solve exactly for the wavefunction
describing the state of the ion.
One can thus obtain an expression for the population remaining
in the ground state as a function of time.  An average of this
population over an ensemble of realizations of the classical
perturbing field (assumed to have Gaussian statistics) yields
simple expressions for the time dependent behavior of the depopulation
both for short and for long time limits.

In the interaction picture the Hamiltonian describing 
the interaction of a single ion of
mass $M$
with a {\em classical} electric field $E(t)$ is given by the
following formula
\begin{eqnarray}
\hat{H}&=&-e E(t)\hat{x} \nonumber \\
&=&i\hbar\left[u(t)\hat{a}^\dagger-u^{*}(t)\hat{a}\right]
\end{eqnarray}
where $\hat{x}$ is the operator for the position of the ion
and $u(t)$ $=$ $ie E(t)\exp(i\omega_{0}t)/\sqrt{2M\hbar\omega_{0}}$,
$e$ being the ion charge and $\hbar$ Planck's
constant divided by $2\pi$.  I
am considering only the motion of the ion along the weak
confinement axis of an anisotropic trap, of the sort suitable
for quantum computation; thus $E(t)$ represents the
component of the electric field along that axis.
The harmonic binding potential is characterized by 
the angular frequency $\omega_{0}$. 
The operators $\hat{a}$ ($\hat{a}^\dagger$) are the zero time annihilation (creation)
operators for the harmonic motion of the ion in the harmonic well.

The dynamics of such a driven quantum harmonic oscillator can be
solved exactly \cite{Glauber:63,MandelWolf}.  The wave function
at some instant $t$ is related to the initial wave function by
the following expression
\begin{equation}
|\psi(t)\rangle=\exp\left[i\phi(t)\right]\hat{D} \left[v(t)\right] 
|\psi(0)\rangle,
\label{darwin}
\end{equation}
where $\hat{D}[v]=\exp\left[v\hat{a}^\dagger-v^{*}\hat{a}\right] $ is the
displacement operator for coherent states of a harmonic
oscillator, $\phi(t)$ is a phase factor (which turns out to be
unimportant for the current problem) and the amplitude
$v(t)$ is given by the formula
\begin{eqnarray}
v(t)&\equiv&\int^{t}_{0}u(t^{\prime})dt^{\prime} \nonumber \\
&=&\frac{i e}{\sqrt{2M\hbar\omega_{0}}}\int^{t}_{0}E(t^{\prime})
\exp(i\omega_{0}t^{\prime}) dt^{\prime} .
\label{vee}
\end{eqnarray}

As a figure of merit, let us introduce an average fidelity of the 
ground state, defined by the following formula:  
\begin{equation}
F(t)=\langle \left| 
\langle\psi(t) |\psi(0)\rangle_{q}\right|^{2}\rangle_{f},
\end{equation}
where $\langle\ldots\rangle_{q}$ denotes a 
quantum mechanical average, and $\langle\ldots\rangle_{f}$ an average 
over an ensemble of realizations of the classical random field $E(t)$.
If one assumes that the ion is initially in the ground state, then its 
state evolves into a coherent state of amplitude $v(t)$.  Thus the 
probability of remaining in the ground state can be found in closed 
form, and the fidelity of the ground state is given by the formula
\begin{equation}
F(t)=\langle \exp\left[-\left|v(t)\right|^{2}\right] \rangle_{f}.
\end{equation}
If one assumes Gaussian statistics for the classical random field $E(t)$
\cite{Goodman}
(which can be justified by assuming that the field is due
to many uncorrelated random sources, and then invoking the central 
limit theorem), the average over the field ensemble can be 
determined by performing an integration using the appropriate
probability distribution.  The result is as follows:
\begin{eqnarray}
F(t)&=&\left[ 1+2\langle \left|v(t)\right|^{2} \rangle_{f}\right.\nonumber \\
&+&\left.
\left(\langle \left|v(t)\right|^{2} \rangle_{f}\right)^{2}
-\left| \langle v(t) v(t) \rangle_{f}\right|^{2}\right]^{-1/2} .
\label{fa}
\end{eqnarray}

An alternative measure of the heating is the mean
excitation number, defined by
\begin{equation}
\bar{n}(t)=\langle\left[\langle\psi(t)|\hat{a}^\dagger\hat{a}
|\psi(t)\rangle_{q}\right]\rangle_{f}.
\end{equation}
Using eq.(\ref{darwin}) and eq.(11.3-13) of ref. \cite{MandelWolf},
this can be rewritten as
\begin{eqnarray}
\bar{n}&=&\langle\left[\langle0|\hat{n}+v^{*}(t)\hat{a}+\hat{a}^\dagger 
v(t)+\left|v(t)\right|^{2}|0\rangle_{q}\right]\rangle_{f} \nonumber \\
&=&\langle\left|v(t)\right|^{2}\rangle_{f}.
\label{quad}
\end{eqnarray}

The correlation functions appearing in eqs.(\ref{fa}) and (\ref{quad}) can be
evaluated using eq.(\ref{vee}).  Using the symmetry property
of the autocorrelation function, one obtains:
\begin{eqnarray}
\langle \left| v(t) \right|^{2} 
\rangle_{f}&=&\Omega^{2}t^{2} \int^{1}_{0} 
(1-x)\gamma_{E}(x t)\cos\left(x\omega_{0}t\right)dx
\label{cc}\\
\langle v(t)^{2} 
\rangle_{f}&=&\frac{\Omega^{2}t}{\omega_{0}}\exp(i\omega_{0}t) \nonumber \\
&\times&\int^{1}_{0} 
\gamma_{E}(x t)\sin\left[(1-x)\omega_{0}t\right]dx
\label{ac}
\end{eqnarray}
where $\gamma_{E}(\tau)=\langle E(t+\tau/2)E(t-\tau/2)\rangle_{f}/
\langle E(t)^{2}\rangle_{f}$ is the degree of correlation
of the field $E(t)$ (which is real) and the characteristic 
transition rate $\Omega$ is given by
$\Omega=\sqrt{e^{2}\langle E(t)^{2}\rangle_{f}/M\hbar\omega_{0}}$.
Since I have implicitly assumed that $E(t)$ is stationary,
$\langle E(t)^{2} \rangle_{f}$ is independent of
time.

For an exponential degree of correlation given by
$\gamma_{E}(\tau)=\exp(-\left|\tau\right|/T)$, the
integrals given in eqs.(\ref{cc}) and (\ref{ac})
can be evaluated in closed form:
\begin{eqnarray}
&&\langle \left|v(t)\right|^{2}\rangle_{f}=\bar{n}(t)= \nonumber \\
&&\,\,\,\,\,\,\,\frac{T}{\tau_{1}}
\left[\exp(-t/T)\cos\left(\omega_{0}t+2\phi\right)
-\cos\left(2\phi\right)+t/T\right] ,
\label{cccf} \\
&&\langle v(t)^{2} \rangle_{f}=\nonumber \\
&&\,\,\,\,\,\,\,\frac{T}{\tau_{1}}\exp\left(i\omega_{0}t\right)
\left[\exp(-t/T)\sin\left(\phi\right)
+\sin\left(\omega_{0}t-\phi\right)\right],
\label{accf}
\end{eqnarray}
where $\tan\phi=\omega_{0}T$ and the heating time $\tau_{1}$ is
given by the formula
\begin{equation}
\frac{1}{\tau_{1}}=
\left(\frac{e^{2}\langle E(t)^{2}\rangle_{f}}{M\hbar\omega_{0}}\right)
\frac{T}{1+\omega_{0}^{2}T^{2}}.
\end{equation}
Examples of the fidelity calculated using
these results are shown in Fig.1.  These show that revivals of the
ground state populations can occur when the heating field has both
a low amplitude and a long coherence time.

\begin{figure}[!ht]
\begin{center}
\epsfxsize=8.5cm  % \columnwidth
\epsfbox{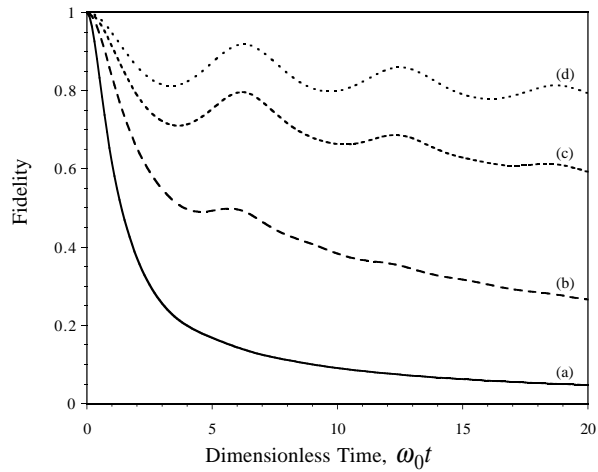}
\end{center}
\caption{The fidelity of the ground state as a function of time,
illustrating the results given in eqs.(\ref{fa}), (\ref{cccf}) and 
(\ref{accf}).  The parameters used were as follows:
curve (a) $\omega_{0}T$=1, $\omega_{0}\tau_{1}$= 1; 
curve (b) $\omega_{0}T$=1, $\omega_{0}\tau_{1}$= 8.5; 
curve (c) $\omega_{0}T$=1, $\omega_{0}\tau_{1}$= 41; 
curve (d) $\omega_{0}T$=1, $\omega_{0}\tau_{1}$= 128.5.
}
\label{fig:fidelity}
\end{figure}

%short time, small bandwidth %
Two limiting cases are of interest, namely the behavior
for short times and for very long times. 
For short times (i.e. $t\ll T,1/\omega_{0}$)
the mean excitation number and fidelity are given by
\begin{equation}
\bar{n}(t)\approx 
1-F(t)=\left(\frac{1+\omega_{0}^{2}T^{2}}{2T\tau_{1}}\right)t^{2}
+O\left(t^{3}\right)
\end{equation}
This result, i.e. that for short times the
``decay'' of the ground state population is non-exponential,
allows the 
possibility of maintaining the ion in its ground state via the {\em 
quantum Zeno effect} \cite{Zurek:84,Itano:90}.  If repeated 
measurements of the
ground state population can be made on time scales much shorter than 
$1/\omega_{0}$, then the ion will in principle
remain in the ground state for a much
longer time. However, as the quadratic behaviour persists for a short
time only (see Fig.1), such a technique would be very difficult to 
implement.  Furthermore, by invoking the time-energy uncertainty 
principle, one can see that measurements carried out on time scales
much less than $1/\omega_{0}$ will result in excitation of higher
number states.  Hence it is unlikely that the Zeno effect will be
a practical method of error correction in these circumstances.

One can find asymptotic expressions for
$F(t)$ and $\bar{n}(t)$in the long time limit  $t\gg T,1/\omega_{0}$:
\begin{eqnarray}
\bar{n}(t)&\sim&\frac{1}{\tau_{1}}(t-t_{0})+O\left(\frac{1}{t^{3}}\right) \\
F(t)&\sim& 
\frac{\tau_{1}}{t}-\frac{\tau_{1}^{2}(1-t_{0}/\tau_{1})}{t^{2}}+
O\left(\frac{1}{t^{3}}\right)
\label{flong}
\end{eqnarray}
where $t_{0}=T(1-\omega_{0}^{2}T^{2})/(1+\omega_{0}^{2}T^{2})$.
These results are in qualitative agreement with that obtained
earlier by Lamoreaux \cite{Lamoreaux:97} using a perturbative
density matrix approach.

% thermal fields %%%%
In order to gain some insight into the order of
magnitude of the heating effect, one can calculate
a formula for the value of $\tau_{1}$ when the
ambient electric field $E(t)$ is thermal, assuming
a correlation time $T\approx\hbar/K_{B}\Theta \ll 1/\omega_{0}$,
(where $K_{B}$ is Boltzmann's constant and $\Theta$ is the
temperature):
\begin{equation}
\tau_{1}\approx\frac{3 c\epsilon_{0}\omega_{0}MK_{B}}
{e^{2}\sigma \Theta^{3}}
\end{equation}
where
$\sigma$ is the  Stefan-Boltzmann constant and 
$\epsilon_{0}$ is the permittivity of free space. 
Using data from the mercury ion experiment described in
ref.\cite{Diedrich:89} (i.e. $M=3.29 \times 10^{-25}$ kg, 
$\omega_{0}=(2\pi) 4.66$MHz),
a heating time  $\tau_{1}$=135 ms implies an effective
temperature of $T$= 4.6 K.  This result should be treated 
with caution: the assumption that the random ambient
field in the vicinity of the ion is thermal is not justified.

% multiple ions
Let us now consider the case of multiple ions confined in a
linear configuration.  Because the ions are interacting via
the Coulomb force, their motion will be strongly coupled.  The
small amplitude fluctuations are best described in terms of 
normal modes, each of which can be treated as an independent
harmonic oscillator\cite{James:98}.  If
there are $N$ ions in the trap, there will be a total of $N$
such modes.  I will
number these modes in order of increasing resonance frequency,
the lowest ($p=1$) mode being the center of mass mode,
in which the ions oscillate as if rigidly clamped together. 
In the quantum mechanical description,
each mode is characterized by creation and annihilation operators
$\hat{a}^\dagger_{p}$ and $\hat{a}_{p}$ (where $p=1,\ldots N$).  The Hamiltonian
in this case is given by the expression
\begin{eqnarray}
\hat{H}&=&-e\sum^{N}_{n =1}E_{n }(t)\hat{x}_{n }(t) \nonumber \\
&=&i\hbar\sum^{N}_{p=1}\left[u_{p}(t)\hat{a}^\dagger_{p}-u_{p}^{*}(t)\hat{a}_{p}\right]
\end{eqnarray}
where $E_{n}(t)$ is the electromagnetic field at the $n$-th ion of 
the string of ions, $\hat{x}_{n}(t)$ is the operator for the displacement 
of the $n-th$ ion from its equilibrium position and
\begin{equation}
u_{p}(t)=\frac{i e}{\sqrt{2M\hbar\omega_{0}\sqrt{\mu_{p}}}}
\sum_{n =1}^{N}E_{n }(t)b^{(p)}_{n }
\exp\left(i\sqrt{\mu_{p}}\omega_{0}t\right).\label{up}
\end{equation}
In eq.(\ref{up}), $b^{(p)}_{n}$ is the $n$-th element of the $p$-th
normalized eigenvector of the ion coupling matrix \cite{James:98},
$\mu_{p}$ being its eigenvalue.  Again the evolution of the state
of the ions can be solved exactly:
\begin{equation}
|\psi(t)\rangle=
\exp\left[i\Phi(t)\right]\prod^{N}_{p=1}\hat{D}_{p} \left[v_{p}(t)\right] 
|\psi(0)\rangle,
\end{equation}
where $\hat{D}_{p}[v_{p}]=\exp\left[v_{p}\hat{a}^\dagger_{p}-v_{p}^{*}\hat{a}_{p}\right]$
is the displacement operator for the $p$-th mode, and 
\begin{equation}
v_{p}(t)=\int^{t}_{0}u_{p}(t^{\prime})dt^{\prime} .
\end{equation}
As before, one can find a closed form expression for the fidelity of 
the ground state of the string of ions. For 
multiple ions, the mean excitation number is given by a  formula analogous to 
eq.(16), where the characteristic decay time is given by:
\begin{equation}
\tau_{N}=\tau_{1}\left[\sum_{p=1}^{N}\sum_{m,n =1}^{N}
\frac{b^{(p)}_{m}b^{(p)}_{n }}{\sqrt{\mu_{p}}}\gamma_{m n}\right]^{-1},
\end{equation}
$\gamma_{m n}$ being the degree of coherence of the field at the 
positions of ions $n$ and $m$ evaluated for zero time delay
\cite{MandelWolf} (the field having been assumed to be
cross-spectrally pure \cite{Mandel,Wolf}), and the coherence
time $T$ has been assumed to be much less than $1/\omega_{0}$. 
In the coherent limit ($\gamma_{m n}=1$), the formula for $\tau_{N}$ 
reduces to the following simple expression
\begin{equation}
\tau_{N}=\frac{\tau_{1}}{N},
\end{equation}
while in the incoherent limit ($\gamma_{m n}=\delta_{m,n}$), $\tau_{N}$
is given by the formula
\begin{equation}
\tau_{N}=\tau_{1}\left[\sum_{p=1}^{N}
\frac{1}{\sqrt{\mu_{p}}}\right]^{-1}.
\label{teenincoh}
\end{equation}
The sum can be worked out from the eigenvalues $\mu_{p}$, which
must be determined numerically in general.  The results are shown in
fig.2.  The separation of ions is generally of the order of a 
few tens of microns.  The coherence length 
of thermal light is $\ell_{coh}\approx\hbar c /K_{B}\Theta=2.3{\rm 
mm}/\Theta$;
at low temperatures it is much longer than the ion separation.  
Therefore 
it is the coherent limit that is the important one,
at least for small numbers of ions.

\begin{figure}[!ht]
\begin{center}
\epsfxsize=8.5cm  % \columnwidth
\epsfbox{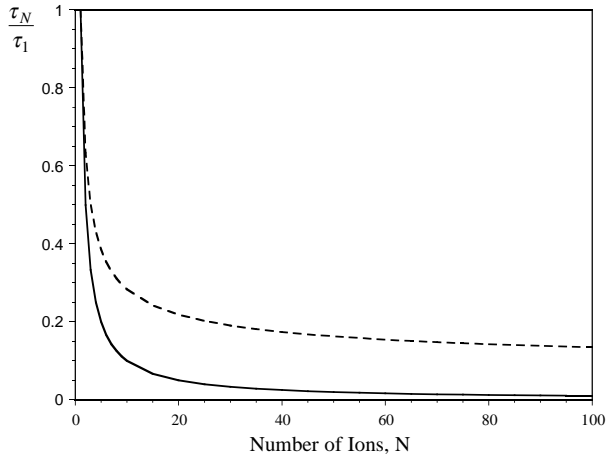}
\end{center}
\caption{The heating time of the ground state for different
numbers of ions, in the spatially coherent (plain curve) and 
spatially incoherent (dashed 
curve) limits for the ambient electric field.}
\label{fig:heatingtimes}
\end{figure}

One can also consider the mean excitation numbers of the different  
modes.  These quantities are given by expressions
analogous to eq.(\ref{quad}), with the heating
time of the $p$-th mode, when there are $N$ ions in the string,
given by
\begin{equation}
\tau_{N,p}=\frac{\sqrt{\mu_{p}}}
{{\displaystyle \sum_{m,n =1}^{N}
b^{(p)}_{m}b^{(p)}_{n }\gamma_{m n}}}\tau_{1}.
\end{equation}
In the coherent and incoherent limits, the
heating times of the different modes are given
\begin{eqnarray}
\tau_{N,p}&=&\left\{
\begin{array}{ll}
\tau_{1}/N&p=1\\
\infty&p>1 .
\end{array}\right.\,\,\,\mbox{coherent}\\
\tau_{N,p}&=&\sqrt{\mu_{p}}\tau_{1}\,\,\,\mbox{incoherent}.
\end{eqnarray}
The heating times for the coherent case
 is potentially a very important result.  Only
the lowest ($p=1$) centre of mass mode will be heated up
by spatially coherent fields.  Thus the state of the ion
oscillations is given by:
\begin{equation}
|\psi(t)\rangle=\exp i\phi(t)\left\{|v_{1}(t)\rangle_{1}\otimes
|0\rangle_{2}\otimes |0\rangle_{3}\otimes
\ldots\otimes|0\rangle_{N}\right\} ,
\end{equation}
the modes other than the center of mass mode remaining in their
ground states.  It should therefore be possible to perform quantum 
logic operations without degradation due to heating, simply by utilizing 
these higher modes as the quantum information bus 
rather than the center of mass mode \cite{Wineland:98}.

The author would like to thank Albert Petschek for reading various
versions of the
manuscript and contributing many useful comments, and
Mark Gulley, Richard Hughes, 
Steve Lamoreaux, Sara Schneider, Dave 
Wineland and Andrew White for other useful conversations 
and correspondence.
This work was funded by the National Security Agency.

\end{document}